\documentclass[aps,twocolumn,floats,superscriptaddress,showpacs]{revtex4}
\usepackage{amsmath,bm}%
\usepackage{graphicx}%

\begin{document}
\title{Application of Information Theory in Nuclear Liquid Gas Phase Transition}
\author{Y. G. Ma}
\thanks{Present Address: Cyclotron Institute, Texas A \& M
University, College Station, Texas 77801}

 %\email{ygma@sinr.ac.cn}

\affiliation{China Center of Advanced Science and Technology
(World Laboratory),
 P. O. Box 8730, Beijing 100080, CHINA}
\affiliation{Shanghai Institute of Nuclear Research, Chinese
Academy of Sciences, P.O. Box 800-204, Shanghai 201800, CHINA}
%\footnotemark \footnotetext{Mailing address.}
%\date{\today}
\date{ PHYS. REV. LETT. {\bf 83} (1999) 3617-3620}

\begin{abstract}
Information entropy and Zipf's law in the field of information theory have
been  used  for studying the disassembly of nuclei in the framework of the isospin
dependent lattice gas model and molecular dynamical model. We found that
the information entropy in the event space is maximum at the phase
transition point  and  the mass  of the cluster show
exactly inversely to its rank, i.e.  Zipf's law appears. Both novel
criteria are useful in searching the nuclear liquid gas phase transition
experimentally and theoretically.
\end{abstract}
\maketitle
%\narrowtext

Hot nuclei can be formed in energetic heavy ion collisions (HIC)
and may be highly excited. They  deexcite by different decay
modes, such as multifragmentation. Experimentally, this kind of
multifragment emission was observed to evolve with beam energy
(excitation energy, or nuclear temperature, ...). Multiplicity,
$N_{imf}$, of intermediate mass fragment (IMF)
 rises with the beam energy, reaches a maximum,
and finally falls  to lower value. This phenomenon of the
rise and fall of $N_{imf}$ may be related to the liquid gas
phase transition in nuclear matter \cite{Maprc95}.
The onset of multifragmentation probably indicates
the coexistence of liquid and gas phases. The mass distribution
of IMF distribution can be expressed as  power law with parameter
 $\tau$. The minimum of  $\tau$,
$\tau _{min}$, occurs when the liquid gas phase transition takes
place \cite{Fisher}. However,  $\tau_{min}$ can also reveal at
supercritical densities along the Kert\'esz line \cite{Camp97}
and at some subcritical densities at lower temperature \cite{Jpan96}.
So it is not possible to determine the phase transition
only from $N_{imf}$ and $\tau _{min}$.

On the other hand,  experimentalists  measured  the nuclear caloric
curves, i.e. the relationship between  nuclear temperature and the excitation
energy. He-Li isotopic temperature from Albergo thermometer
\cite{Albergo}  for projectile-like Au spectators seems to
exhibit a temperature plateau in the excitation energy range
of 3 to 10 MeV/u \cite{Aladin}. This plateau was taken as an
indication for  a first order nuclear liquid gas phase
transition. However, due to the changing mass of Au
spectators with excitation energy and the side-feeding effect to measured
He-Li isotopic temperature, this conclusion is
questionable \cite{Natowitz}. Nuclear caloric curves were also
surveyed by several groups \cite{Maplb97}. However,
unfortunately,  the sharp signature of liquid gas phase transition
in macroscopic systems may be  smoothed and blurred due to
the small numbers of nucleons in nuclei, and/or the difficulty
to perform a direct comparison between the measured "apparent"
temperature and the "real" temperature interferes  obtaining the
 real nuclear caloric curve. These factors hamper the reaching of a
 definite  conclusion on liquid gas phase transition in nuclei.

The extraction of  critical exponents and the study of critical
behavior in  finite-size systems were attempted in \cite{Eos2}
and were followed by controversial debates \cite{Bauer}. In
this context, it is  necessary and meaningful to search for some
novel signatures to characterize the nuclear liquid gas phase
transition in order to guide the experimental  analysis and
theoretical predictions.

In this Letter, we will introduce information entropy \cite{Denb85},
$H$,  and Zipf's law \cite{Crystal} into the diagnosis of nuclear
liquid gas phase transition. The information entropy was
defined by Shannon in information theory.
Originally, it measures the
"amount of information" which is contained in messages sent
along a transmission line.
It can be expressed as follows,
\begin{equation}
 H = -\sum_{i} {p_i ln(p_i)}.
\end{equation}
where ${p_i}$ is a normalized probability, and $\sum_i p_i = 1$.
Jaynes proposed that a very general technique for discovering the
least biased distribution of the $p_i$ consists in the
maximization of the Shannon $H$ entropy, subject to whatever
constraints on $p_i$ are appropriate to the particular situation.
The maximization of $H$ was thus put forward as a general
principle of statistical inference - one which could be applied to
a wide variety of problems in economics, engineering and many
other fields, such as quantum phenomena \cite{Denb85}.  In high
energy hadron collisions, multiparticle production proceeds on the
maximum stochasticity, i.e.  they should obey the maximum entropy
principle. This kind of stochasticity can be also quantified via
the information entropy which  has been shown to be a good tool to
measure chaoticity in hadron decaying branching process
\cite{Brog}. In different physical conditions, information entropy
can be expressed with different stochastic variables. In this work
on HIC, we define $p_i$ as the event probability of having $"i"$
particles produced, i.e. $\{p_i\}$ is the normalized probability
distribution of total multiplicity, the  sum is taken over whole
$\{p_i\}$ distribution. This emphasis is on the event space rather
than the phase space. As shown below  this
 kind of information entropy \cite{Comments} can be taken as a
method to determine nuclear liquid gas phase transition.

Zipf's law \cite{Crystal} has been known as a statistical
phenomenon concerning the relation between English words and their
frequency in literature in the field of linguistics. The law states
that, when we list the words in the order of decreasing population,
the frequency of a word is inversely proportional to its rank
 \cite{Crystal}. This relation was
found not only in linguistics but also in other fields of sciences.
For instance, the law appeared in distributions of populations in
cities, distributions of income of corporations,
distributions of areas of lakes and
cluster-size distribution in percolation process \cite{Watanabe}.
In this Letter, Zipf's law will be tested for the fragment mass
distribution and it is evidenced to be a factor to characterize
the phase transition.

The tools we will use here are the isospin dependent lattice gas
model (LGM) and  molecular dynamical model (MD). The lattice gas
model was developed to describe the liquid-gas phase transition
for atomic system by Lee and Yang \cite{Yang52}. The same
model has already been applied to nuclear physics for isospin
symmetrical systems in the grandcanonical ensemble \cite{Biro86}
with a sampling of the canonical ensemble
\cite{Camp97,Jpan96,Mull97,Jpan95,Jpan98,Gulm98,Ma99}, and also
for isospin asymmetrical nuclear matter in the mean field
approximation \cite{Sray97}. In addition, a classical molecular
dynamical model is used to compare its results with the results
of lattice gas model. Here we will make a brief description for
the models.

In the lattice gas  model, $A$ (= $N + Z$) nucleons with an
occupation number $s$ which is defined $s$ = 1 (-1) for a proton
(neutron) or $s$ = 0 for a vacancy, are placed on the $L$ sites of
lattice. Nucleons in the nearest neighboring sites
interact with an energy $\epsilon_{s_i s_j}$. The hamiltonian
is written as
%\begin{equation}
$E = \sum_{i=1}^{A} \frac{P_i^2}{2m} -
\sum_{i < j} \epsilon_{s_i s_j}s_i s_j $.
%\end{equation}
The interaction constant $\epsilon_{s_i s_j}$ is chosen to
be isospin dependent and be fixed to reproduce the binding
energy of the nuclei \cite{Jpan98}:
\begin{eqnarray}
 \epsilon_{nn} \ &=&\ \epsilon_{pp} \ = \ 0. MeV \nonumber , \\
 \epsilon_{pn} \ &=&\ - 5.33 MeV.
\end{eqnarray}
A three-dimension cubic lattice with $L$ sites is used. The
freeze-out density of disassembling system is assumed to be
 $\rho_f$ = $\frac{A}{L} \rho_0$, where $\rho_0$ is the normal
 nuclear density. The disassembly of the system
is to be calculated at $\rho_f$, beyond which nucleons are too far
apart to interact.  Nucleons are put into lattice by Monte Carlo
Metropolis sampling. Once the nucleons have been placed we also
ascribe to each of them a momentum by Monte Carlo samplings of
Maxwell-Boltzmann distribution.

Once this is done the LGM immediately gives the cluster
distribution using the rule that two nucleons are part of the
same cluster if $P_r^2/2\mu - \epsilon_{s_i s_j}s_i
s_j < 0 $. This method is similar to the Coniglio-Klein's
prescription \cite{Coni80} in condensed matter physics and
was shown to be
valid in LGM \cite{Camp97,Jpan96,Jpan95,Gulm98}. To calculate
clusters using MD we propagate the particles from the initial
configuration for a long time under the influence of the
chosen force. The form of the force is chosen to be also isospin
dependent in order to compare with the results of LGM. The
potential for unlike-nucleons is
\begin{widetext}
\begin{eqnarray}
 v_{\rm n p}(r) (r/r_0<a)\ &=&\ C\left[B(r_0/r)^p-(r_0/r)^q\right]\nonumber
    exp({\frac{1}{(r/r_0)-a}}), \\
v_{\rm  n p}(r) (r/r_0>a)\ &=&\ 0.
\label{pot}
\end{eqnarray}
\end{widetext}
where $r_0 = 1.842 fm$ is the distance between the centers of two
adjacent cubes. The parameters of the potentials are $p$ = 2, $q$
= 1, $a$ = 1.3, $B$ = 0.924, and $C$ = 1966 MeV. With these
parameters the potential is minimum at $r_0$ with the value -5.33
MeV, is zero when the nucleons are more than 1.3$r_0$ apart and
becomes strongly repulsive when $r$ is significantly less than
$r_0$. The potential for like-nucleons is written as
\begin{eqnarray}
v_{\rm p p}(r) (  r < r_0 )\ &=&\  v_{\rm n p}(r)- v_{\rm n p}(r_0)\nonumber , \\
v_{\rm p p}(r) ( r > r_0 )\ &=&\ 0.
\end{eqnarray}
The system evolves with the above potential. At asymptotic times
the clusters are easily recognized. Observables based on the
cluster distribution in the both models can now be compared. In
the case of proton-proton interactions, the Coulomb interaction
can also be added separately and it can be compared with the case
without Coulomb effects.

In this Letter we choose the medium size nuclei $^{129}$Xe  as
an example. In most cases, $\rho_f$ is chosen to be 0.38 $\rho_0$,
since the experimental  data can be  best fitted by $\rho_f$
between 0.3$\rho_0$ and 0.4$\rho_0$ in  previous LGM calculations
\cite{Jpan95,Beau96}, which corresponds to  $7^3$ cubic lattice.
  In addition,  0.18$\rho_0$, corresponding
to $9^3$ cubic lattice and 0.60$\rho_0$, corresponding to $6^3$
cubic lattice of $\rho_f$ are also taken to compare and check the
results with different $\rho_f$ values in the LGM case. 1000 events were
simulated for each combination of $T$ and $\rho_f$ which ensures enough
statistics for results.

In order to check the phase transition behavior in  LGM and MD,
we will first show the results of some physical observables, namely
the effective power-law parameter, $\tau$, the second moment of the
cluster distribution, $S_2$, and the multiplicity of intermediate
mass fragments, $N_{imf}$, for the disassembly of $^{129}$Xe:
fig.1.   These observables were shown to be good indicators of a
liquid gas phase transition, as shown in Ref. \cite{Jpan98}.
The extreme values of $\tau$, $N_{imf}$ and $S_2$ occur at the
same temperature, indicating of the onset of the phase
transition, for each calculation case. For the LGM case, the
phase transition temperature increases with the freeze-out
density; for the MD case,  a slight small transition temperature
is obtained when Coulomb force is ignored. It
becomes much lower in case of Coulomb interaction
due its long range repulsion. Similar phenomenon has been
explored in a previous study \cite{Jpan98}. However,
the aim of this Letter  is to take the above transition
temperatures as references to search for novel signatures of
liquid gas phase transition.
\begin{center}
\begin{figure}
\includegraphics[scale=0.40]{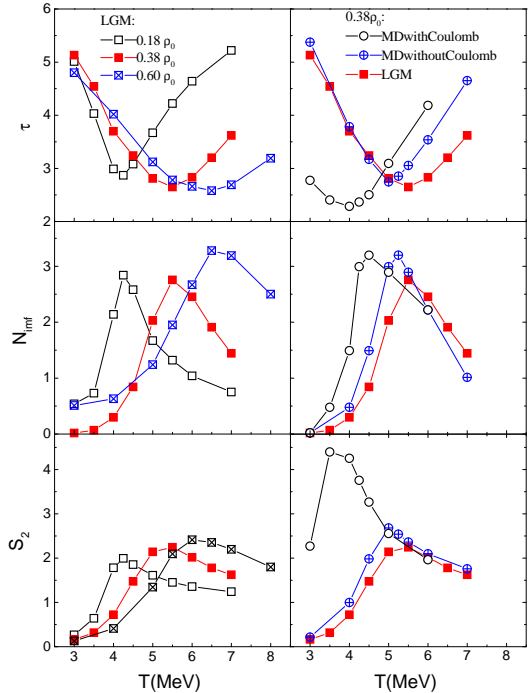}
\caption{\footnotesize The effective power-law parameter, $\tau$,
the second moment of the cluster distribution, $S_2$, and the
multiplicity of intermediate mass fragments, $N_{imf}$ as a
function of temperature for the disassembly of $^{129}$Xe. Left
panel is the LGM results with different $\rho_f$ and right panel
is the comparison of MD to LGM with 0.38$\rho_0$. The symbols
 are illustrated in figure. }
\end{figure}
\end{center}

Fig.2 shows the information entropy for disassembly of Xe. The
information entropy exhibits a rise  and fall with temperature,
which is similar to the behaviors of $N_{imf}$ and $S_2$.
The temperatures extracted from the peak values of $H$  are
consistent with the transition  temperatures  in Fig.1,
indicating that information entropy ought to be a good diagnosis
of phase transition. Physically, the maximum
of $H$ reflects the largest fluctuation  of the multiplicity
probability distribution in the phase transition point. In
this case it is the most difficult to predict how
many clusters will be produced in each event, i.e. the
disorder (entropy) of information is the largest.
Generally speaking, the larger the dispersal of multiplicity
probability distribution, the higher the information entropy
and then the disorder of system in the event topology.  One
should make a careful distinction between  this information
entropy, on the one hand, and the original thermodynamic entropy,
on the other hand \cite{Denb85,Comments}. The latter
generally illustrates the heat disorder in momentum space
rather than event space and it always increases with
temperature.
\begin{center}
\begin{figure}
\includegraphics[scale=0.40]{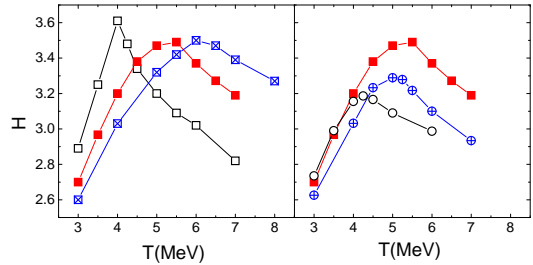}
\caption{\footnotesize The same as Fig.1, but for the information
entropy $H$. }
\end{figure}
\end{center}

Next we will present the results for testing Zipf's law
in the fragment distribution. The law states that the
relation between the sizes and their ranks is described by
$A_n = c/n$ (n=1, 2, 3, ...), where $c$ is a constant and $A_n$
is the mass of rank $n$ in a mass list when we arrange the
clusters in the order of decreasing size. In the calculations,
we averaged the masses for each rank in mass lists
of the events. Then we examined the
relation between the masses $A_n$ and their ranks $n$  with
the fit of $A_n \propto n^{-\lambda}$ in
the range of 1 $\leq$ $n$ $\leq$ 10, where $\lambda$ is
the slope parameter. The upper panel of Fig.3 summarizes
such parameter $\lambda$ as a function of temperature for
LGM and MD cases. Clearly, the value of $\lambda$ decreases
with temperature, indicating that the difference of mass
between the different fragment ranks is becoming smaller. When
$\lambda$ $\sim$ 1, the Zipf's law is satisfied:
$A_n \propto n^{-1}$. The temperatures having Zipf's
law for all calculations in Fig.3 are  also consistent
with the respective transition temperatures extracted
from the extreme values of some observables in Fig.1
and 2. Therefore Zipf's law is also a good signal  to
phase transition. From the statistical
point of view, Zipf's law is related to the
critical behavior or self-organized criticality
\cite{Fisher,Stauffer}, which may be a special
state with the maximum information or least effort.

In order to further illustrate that Zipf's law exists at the phase
transition point most probably, we directly fit the
rank-classified cluster distribution with Zipf's law  and extract
the truth of the hypothesis: $\chi^2$ test. The lower panel of
Fig.3 demonstrates the $\chi^2/ndf$ for the $A_n$ - $n$ relations
at different $T$ for different calculation cases. As expected,
there are minima of $\chi^2/ndf$ around the respective transition
temperature, which further support  Zipf's law of the fragment
distribution indicates a  phase transition. All calculations  give
the same conclusions as above.
\begin{center}
\begin{figure}
\includegraphics[scale=0.40]{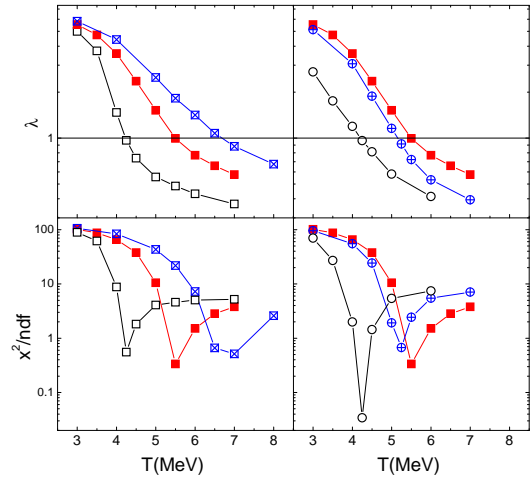}
\caption{\footnotesize  The slope parameter $\lambda$ of the
relation $A_n$ to $n$
 (upper panel) and the $\chi^2/ndf$ with the fit of Zipf's law
(lower panel) as a function of temperature. Left panel is the LGM
results with different $\rho_f$ and right panel is the comparison
of MD to LGM with 0.38$\rho_0$. The symbols are the same as Fig.1.
}
\end{figure}
\end{center}

In addition, we also investigated  larger systems,
such as $A$ = 274, 500 and 830, in the LGM case to see
if the system behaves as expected in $^{129}$Xe.
The results show that  the maximum of information entropy
and Zipf's law behavior still remain at the same phase
transition  temperature as the one extracted from the extreme
values of  $\tau$, $N_{imf}$
and $S_2$.  It illustrates that both novel criteria
are  suitable as signals  of phase transition
in larger $A$ limit.

In conclusion, the information entropy and Zipf's law in
the field of information theory are introduced, for
the first time, into the study of the liquid gas phase
transition of nuclei in the framework of the isospin
dependent lattice gas model and molecular dynamical model
in $\rho$-T plane. At the point of phase transition, the
information entropy of multiplicity distribution is maximum,
which  indicates that the system at this time  has the largest
fluctuation/stochasticity/chaoticity in the event
space. Meanwhile, the cluster mass  show exactly
inversely to its rank, i.e. Zipf's law appears.  Even
thought both criteria are still phenomenological,
we believe that they are simple and
practicable tools to diagnose the nuclear liquid gas phase
transition in experiments and theories.  We are waiting for
 some data analysis on the information entropy and Zipf's
 law in near future.

It is my pleasure to thank Dr. J. Pan and Dr. S. Das Gupta for
providing the LGM and MD codes and Dr. J. P\'eter for reading the
paper. This work was supported  by the China's Distinguished Young
Scholar Fund under Grant No. 19725521, the NSFC under Grant No.
19705012, the STDF of Shanghai under Grant No. 97QA14038, and the
Presidential Foundation of CAS.

{}

\footnotesize
\begin{thebibliography}{}
\bibitem{Maprc95}C.A. Ogilvie {\it et\ al}.,
Phys.\ Rev.\ Lett.\ {\bf 67}, 1214
(1991); M.B. Tsang {\it et\ al}.,  Phys.\ Rev.\ Lett.\
{\bf 71}, 1502 (1993);
Y.G. Ma and W.Q. Shen, Phys.\ Rev.\ C {\bf 51}, 710 (1995).
\bibitem{Fisher}M.E. Fisher, Physics (N.Y.) {\bf 3}, 255 (1967).
\bibitem{Camp97}X. Campi  and H. Krivine,
Nucl.\ Phys.\ A {\bf 620}, 46 (1997).
\bibitem{Jpan96}J. Pan  and S. Das Gupta,
Phys.\ Rev.\ C {\bf 53}, 1319 (1996).
\bibitem{Albergo}S. Albergo {\it et\ al}.,
Nuovo\ Cimento\ A {\bf 89}, 1 (1985).
\bibitem{Aladin}J. Pochodzalla {\it et\ al}.,
Phys.\ Rev.\ Lett.\ {\bf 75}, 1040
(1995).
\bibitem{Natowitz} J.B. Natowitz {\it et\ al}.,
Phys.\ Rev.\ C {\bf 52}, R2322 (1995);
M.B. Tsang {\it et\ al}., Phys.\ Rev.\ Lett.\
{\bf 78}, 3836 (1997); A. Siwek {\it
et\ al}., Phys.\ Rev.\ C {\bf 57}, 2507 (1998).

\bibitem{Maplb97}Y.G. Ma {\it et\ al}.,
Phys.\ Lett.\ B {\bf 390}, 41 (1997);
M.J. Huang {\it et\ al}., Phys.\ Rev.\ Lett.
{\bf 78}, 1648 (1997); R. Wada {\it et\ al}.,
Phys.\ Rev.\ C {\bf 55}, 227 (1997);  J.A. Hauger
{\it et\ al}., Phys.\ Rev.\ Lett.
{\bf 77}, 235 (1997); V. Serfling {\it et\ al}.,
Phys.\ Rev.\ Lett. {\bf 80}, 3928
(1998).

\bibitem{Eos2}M.L. Gilkes {\it et\ al}.,
Phys.\ Rev.\ Lett. {\bf 73}, 1590(1994);
J.B. Elliott {\it et\ al}.,  Phys.\ Lett.\ B
{\bf 381}, 35 (1998); P.F. Mastinu {\it
et\ al}., Phys.\ Rev.\ Lett. {\bf 76}, 2646 (1996);
M.L. Cherry {\it et\ al}., Phys.\
Rev.\ C {\bf 52}, 2652 (1995).
\bibitem{Bauer}J.B. Elliot {\it et\ al}.,
Phys.\ Rev.\ C {\bf 55}, 544 (1997); W.
Bauer and A. Botvina, Phys.\ Rev.\ C {\bf 55},
546 (1997) and references therein;
L.G. Moretto {\it et\ al}., Phys.\ Rev.\ Lett.\
{\bf 79}, 3538 (1997).

\bibitem{Denb85}K.G. Denbigh and  J.S. Denbigh,
{\it Entropy in Relation to Uncomplete Knowledge},
Cambridge University Press, 1995.
\bibitem{Crystal} G.K. Zipf, {\it Human\ Behavior\ and\ the\
Principle\ of\ Least\ Effort}, Addisson-Wesley Press, Cambridge,
MA, 1949; D. Crystal, {\it The Cambridge
Encyclopedia of Language},
Cambridge University Press, Cambridge, 1987, p86.
\bibitem{Brog}P. Brogueira {\it et\ al}.,
Phys.\ Rev.\ D {\bf 53}, 5283 (1996); Zhen
Cao and R.C. Hwa, Phys.\ Rev.\ D {\bf 53}, 6608 (1996).

\bibitem{Comments}The information entropy
Eq.(1) is defined only if all events $i$ are $a\ priori$ equally probable
in equilibrium thermodynamics. However, the equal probablity of events
is nearly impossible in hadron collisions and nuclear fragmentation since
the process is highly dynamical. A theory based on
equilibrium thermodynamics would in any case be only an approximation.
In this context a phenomenological study is probably more useful than
a thermodynamical theory. Of course the entropy defined here is not
strict information entropy. Probably, it is better to call it
"event entropy"  or "multiplicity entropy". However, since the form
of information entropy Eq.(1) has been  well known, we will still  call it
information entropy as already made in hadron collision (see \cite{Brog})
throughout this phenomenological study.

\bibitem{Watanabe}M. Watanabe, Phys.\ Rev.\ E {\bf 53}, 4187 (1996).

\bibitem{Yang52}T.D. Lee and C.N. Yang,
Phys.\ Rev.\ {\bf 87}, 410 (1952).
\bibitem{Biro86}T.S. Biro {\it et\ al}.,
Nucl.\ Phys.\ A {\bf 459}, 692  (1986);
S.K. Samaddar  and J. Richert,
Phys.\ Lett.\ B {\bf 218}, 381 (1989);
Z.\ Phys.\ A {\bf 332}, 443 (1989);
J.M. Carmona {\it et\ al}., Nucl.\ Phys.\ A {\bf 643}, 115 (1998).
\bibitem{Mull97}W.F.J. M\"uller, Phys.\ Rev.\ C {\bf 56}, 2873 (1997).
\bibitem{Jpan95}J. Pan  and S. Das Gupta, Phys.\ Lett.\
B {\bf 344}, 29 (1995);
 Phys.\ Rev.\ C {\bf 51}, 1384 (1995);
 Phys.\ Rev.\ Lett.\ {\bf 80}, 1182 (1998);
 S. Das Gupta {\it et\ al}.,  Nucl.\ Phys.\ A {\bf 621}, 897 (1997).
\bibitem{Jpan98} J. Pan and S. Das Gupta,
Phys.\ Rev.\ C {\bf 57}, 1839 (1998).
\bibitem{Gulm98}F. Gullminelli and P. Chomaz,
Phys.\ Rev.\ Lett.\ {\bf 82}, 1402 (1999).
\bibitem{Ma99}Y.G. Ma {\it et\ al}., Euro.\ Phys.\ J.\ A {\bf 4}, 217 (1999);
J.\ Phys.\  G{\bf 25}, 1559 (1999).
\bibitem{Sray97}S. Ray {\it et\ al}., Phys.\ Lett.\ B {\bf 392}, 7 (1997).
\bibitem{Coni80}A. Coniglio and E. Klein, J.\ Phys.\ A {\bf 13}, 2775 (1980).
\bibitem{Beau96}L. Beaulieu {\it et\ al}.,
Phys.\ Rev.\ C {\bf 54}, R973 (1996).
\bibitem{Stauffer}D. Stauffer, {\it Introduction\ to\
Percolation\ Theory}, Taylor and Francis, London, 1985.
\end{thebibliography}
\end{document}